\documentclass[twocolumn,a4paper,10pt]{article}
\usepackage{amsmath,amsfonts}
\usepackage{epsfig}
\usepackage{times}
\usepackage{epsfig}
\usepackage{enumitem}
\usepackage[small,hang]{caption}
\usepackage{amssymb}
\usepackage{lipsum}
\usepackage{pslatex}
\usepackage{epsfig,soul}
\usepackage{caption,subcaption}
\usepackage{booktabs,multirow}
\usepackage{graphicx,tikz,graphics,epsfig}
\usepackage{url,CJKutf8}
\usepackage{color,xcolor}
\usepackage{multirow}
\usepackage{physics}
\usepackage{natbib}

\newcommand{\RomanNumeralCaps}[1]
    {\MakeUppercase{\romannumeral #1}}

\makeatletter  

\newcounter{numbersec}
\renewcommand{\section}[1]{\par\noindent\stepcounter{numbersec}
	\par
	\vspace{6pt}
	\noindent\textbf{\large   \arabic{numbersec} \hspace*{0.3cm} #1 }
	\par
	\vspace{2pt}
}
\renewcommand{\subsection}[1]{
	\par
	\vspace{6pt}
	\noindent\textbf{#1}
	\par
}
\renewcommand{\subsubsection}[1]{%
	\par
	\vspace{6pt}
	\textbf{#1.}
}

\newcommand{\Abstract}{\par\vspace{6pt}\noindent\textbf{\large Abstract}\par\vspace{2pt}}
\newcommand{\Acknowledgments}{\par\vspace{6pt}\noindent\textbf{\large Acknowledgments }\par\vspace{2pt}}

\newenvironment{References}{
\par\vspace{6pt}\noindent\textbf{\large References}\par\vspace{2pt}
\begin{small}\begin{list}{ }{\itemsep2mm \parsep0mm\labelsep0mm\leftmargin0mm}}
{\end{list}\end{small}}

\makeatother

\title{\vspace*{-12mm}
\LARGE \sc \textbf{  
Minimum-dissipation model for large-eddy simulation using symmetry-preserving discretization in OpenFOAM \\
}}

\author{ \Large \bf \textit{ 
J Sun$^{1}$ and R.W.C.P Verstappen$^{1}$}  \\ \\
\bf  $^{1}$ \textit{Computational and Numerical Mathematics, University of Groningen, The Netherlands} \\
{\it j.sun@rug.nl}
}
\date{}

\begin{document}
\maketitle
\thispagestyle{empty}



\Abstract
The minimum-dissipation model is applied to channel flow up to $Re_\tau = 2000$, flow past a circular cylinder at $Re=3900$, and flow over periodic hills at $Re=10595$. Numerical simulations were performed in OpenFOAM which is based on the finite volume methods. We used both symmetry-preserving and standard second-order accurate discretization methods in OpenFOAM on structured meshes. The results are compared to DNS and experimental data. 

The results of channel flow demonstrate a static QR model performs equally well as the dynamic models while reducing the computational cost. The model constant of $C=0.024$ gives the most accurate prediction, and the contribution of the sub-grid model decreases with the increase of the mesh resolution and becomes very small (less than 0.2 molecular viscosity) if a fine mesh is used. Furthermore, the QR model is able to predict the mean and rms velocity accurately up to $Re_\tau = 2000$ without a wall damping function. The symmetry-preserving discretization outperforms the standard OpenFOAM discretization at $Re_\tau=1000$. The results for the flow over a cylinder show that the mean velocity, drag coefficient, and lift coefficient are in good agreement with the experimental data and the central difference schemes conjugated with the QR model predict better results.
The various comparisons carried out for flows over periodic hills demonstrate the need to use central difference schemes in OpenFOAM in combination with the minimum dissipation model. The best model constant is again $C=0.024$.
The single wind turbine simulation shows that the QR model is capable of predicting accurate results in complex rotating scenarios.


\section{Introduction} 

Minimum-dissipation models aim to describe subfilter turbulent fluxes in large-eddy simulation. The first minimum-dissipation eddy-viscosity model is the QR model proposed by Verstappen (2011 and 2018). 
The QR model has many desirable properties. It is more cost-effective than the dynamic Smagorinsky model, it appropriately switches off in laminar and transitional flows, and it is consistent with the exact subfilter stress tensor on isotropic grids. 
Few studies have investigated the QR model, especially in open-source software. 
In this work, we implement the QR model in OpenFOAM and perform simulations at high-Reynolds-number and in complex geometries, making this the first study of its kind.

Verstappen and Veldman (2003) proposed to describe Navier-Stokes equations such that the symmetries of the differential operators are preserved on a staggered non-uniform grid, i.e., the convective operator is represented by a skew-symmetric matrix and the diffusive operator by a symmetric, positive-definite matrix. Trias et al (2014) generalized this method to unstructured collocated meshes and proposed an approach, based on a fully-conservative regularization of the convective term,  to mitigate the checkerboard spurious modes. Building upon these ideas, Komen et al (2021) developed a conservative symmetry-preserving second-order time-accurate PISO-based pressure-velocity coupling method for solving the incompressible Navier-Stokes equations on unstructured collocated grids. They implemented this approach in OpenFOAM. The code used in the present study is based on that; it is provided by Hopman (2023).

\section{Minimum-dissipation model} 
Large eddy simulation (LES) represents the large turbulent motions directly and models the effect of the small-scale motions with a sub-grid model.  The minimum-dissipation model is based on the invariants of the rate of strain tensor. By confining the sub-grid kinetic energy with Poincaré's inequality, the minimum amount of eddy viscosity needed to counteract the nonlinear production is given by $\nu_e = C_\Delta \overline{|r(v)|} / \overline {q(v)}$, where $q$ and $r$ are the second and third invariant of the rate of strain tensor (the first invariant is $\nabla\cdot v = 0$); $C_\Delta$  depends on the filter length. Note that the Smagorinsky model only depends on $q(v)$, i.e., not on r(v). The resulting eddy viscosity vanishes in any laminar (part of the) flow since $r=0$ in laminar flow. At a no-slip wall $r=0$ as well; hence $\nu_e=0$ at the wall. 

The grid cell average of invariants is approximated by, the mid-point, rule for integration. This gives the QR model 
\begin{equation}
    \label{eq:31}
    \tau -\frac{1}{3} tr(\tau) I = -2\nu_e S(v) = -2 C_\Delta \frac{|r(v)|}{q(v)} S(v)
\end{equation}

\section{Channel flow} 
The numerical investigations of both static and dynamic minimum-dissipation models applied to channel flow are presented, for friction Reynolds numbers up to $Re_\tau = 2000$ (based on the half channel width). The model contribution on different mesh resolutions is studied and the symmetry-preserving discretization is compared with the standard OpenFOAM discretization at $Re_\tau =1000$.  
\subsection{Physical and Numerical Domain}
Fully developed channel flow is homogeneous in the streamwise and spanwise directions, hence periodic boundary conditions are used in these directions. The boundary conditions on the wall are no-slip for velocity, zero pressure gradient, and vanishing eddy viscosity. The mesh distribution is uniform in the streamwise and spanwise directions and stretched in the wall-normal direction (clustered near the walls). The velocity field is initialized using the minimum-dissipation model’s results obtained on a coarser grid after 10000-time steps with the help of OpenFOAM build-in function mapFields. In this way, fewer time steps are needed before starting the averaging process. The time step is chosen so that the Courant-Friedrichs-Lewy number is less than 0.8 in every simulation. Only a few hundred-time steps are required with this method to obtain a fully developed turbulent flow. The bulk velocity and kinetic viscosity are pre-set, friction velocity $u_{\tau}=\sqrt{\tau_{w}/\rho}$, $\tau_w = \nu \partial{u}/\partial {y}$ is calculated from the wall shear stress.  

\subsection{Results and discussion}

\subsubsection{Optimal QR model constant at Re$_\tau$ = 180}
\begin{figure}[b!]
	\begin{center}
	\includegraphics[width=0.8\linewidth]{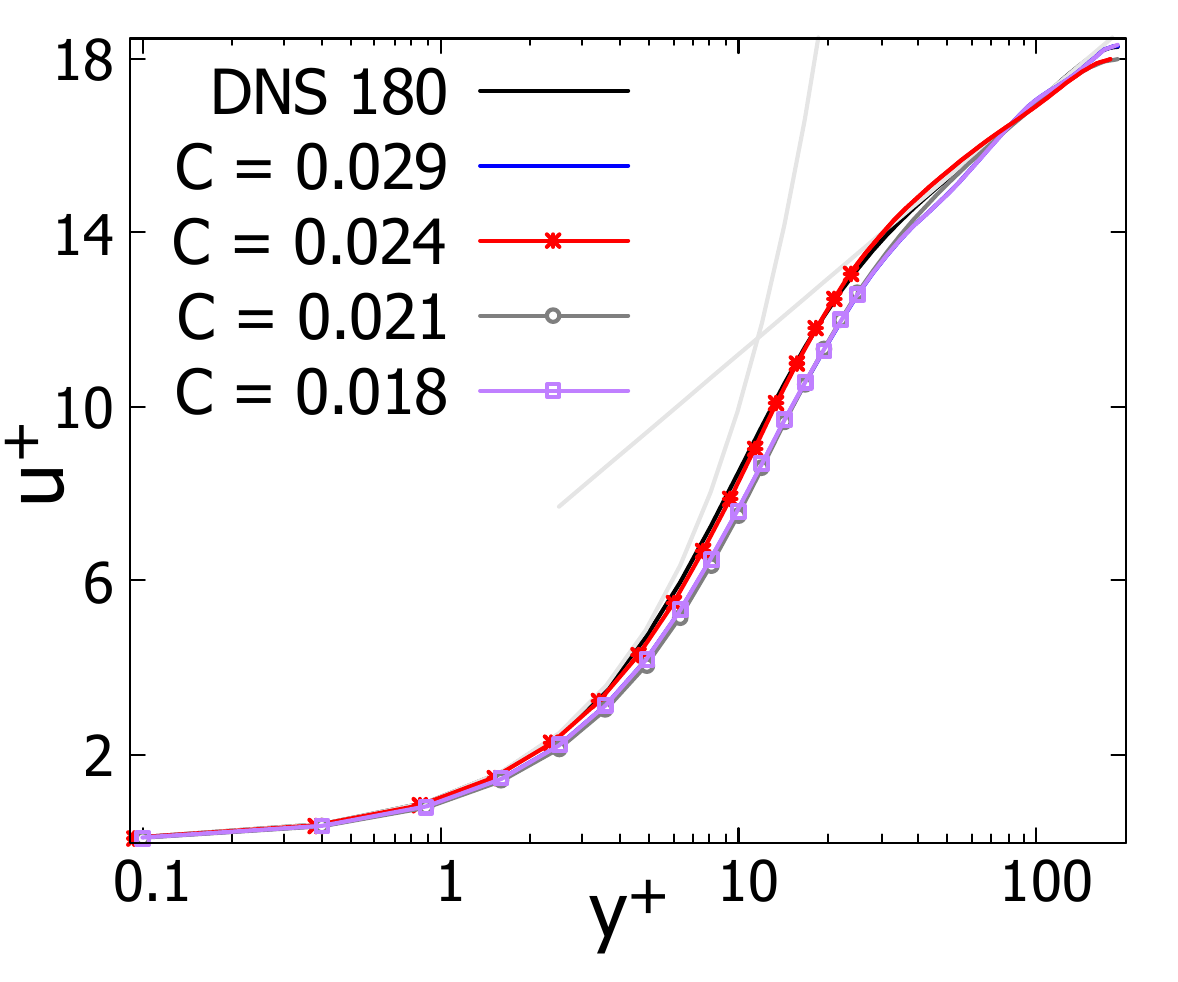}
	\caption{Normalized mean streamwise velocity against wall distance in wall units at $Re_\tau=180$. The computational domain is $4\delta \times 2 \times 2\delta$, the grid resolution is $40\times 50\times 30$. }
        \label{fig:channel2} 
	\end{center}
\end{figure}
In this part, QR models with different model constants are applied to the channel flow at $Re_\tau=180$. Note that the model constant used in the minimum-dissipation model corresponds to the square of the Smagorinsky model constant. The normalized uniform grid spacing is $\Delta x^+=18$ in streamwise direction and $\Delta z^+=12$ in spanwise. The grading expansion in wall-normal direction is around 10, the first normalized wall-normal grid point next to the wall is  $\Delta y_w^+=1.728$. The time step for simulation is $\Delta t^+=36$.

Figure  \ref{fig:channel2} shows the normalized mean streamwise velocity against the wall distance in the wall unit. As we can see from Figure \ref{fig:channel2}, a small value, for instance, C = 0.018, underestimates the mean velocity in the whole channel. This is also the case for C = 0.029. While the medium value of C = 0.024 is precisely in line  with the DNS data. 

The errors quantified with the five measurements show that the model constant C = 0.024 gives the smallest error (mean square error, absolute error, maximum absolute error, and slope error). The value of C = 0.023, however, gives the lowest error if the van Karman constant is used to quantify the error. Note that this measurement considers only the difference in the logarithmic region ($y^+>30$).  

These findings indicate that  C = 0.023 is more accurate in the log wall region. In the near wall region, the optimal minimum-dissipation model constant is C = 0.024. According to the literature, the best value of the Smagorinsky model is between $C_s=0.1$ and $C_s=0.2$.  Thus, the optimal constant of the QR model is found in the range of $C_s^2$. 

\subsubsection{Static QR model in comparison to dynamic models at Re$_\tau$ = 180}
\begin{figure}[ht!]
    \centering
    \includegraphics[width=0.8\linewidth]{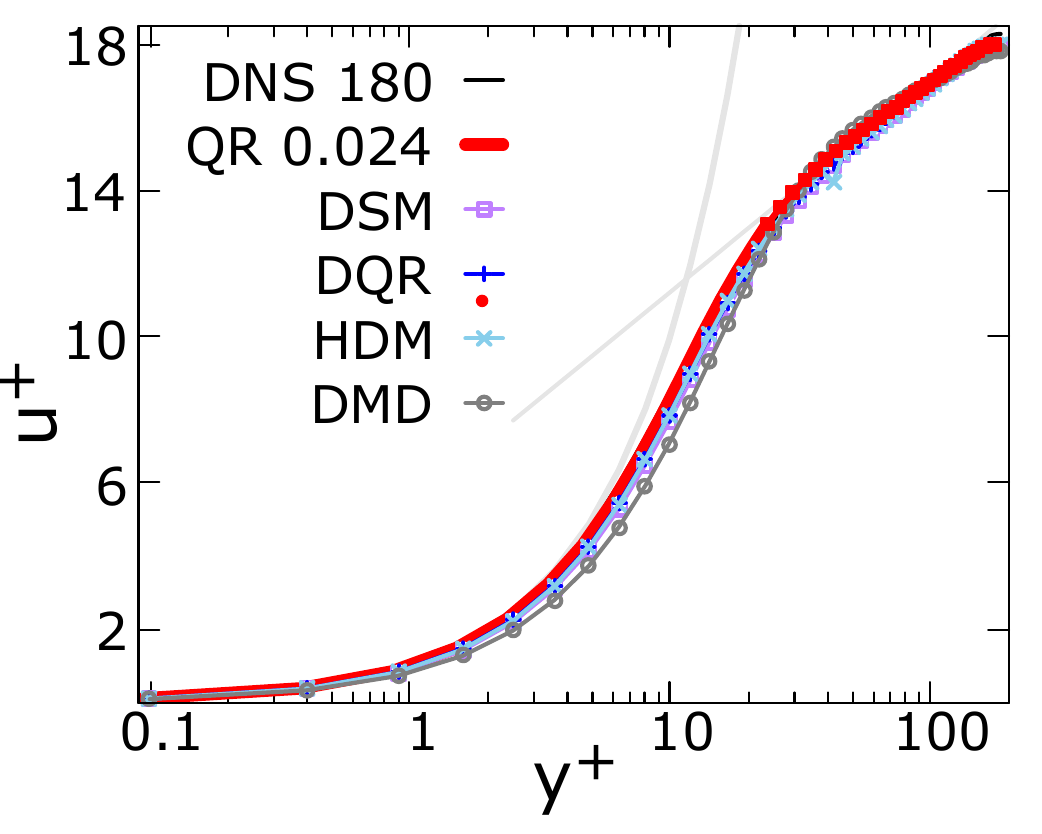}
    \caption{Normalized streamwise velocity $u^+$ against wall distance in wall units $y^+$ at $Re_\tau=180$. Five models are compared to DNS data: QR 0.024: QR model with C=0.024; DQR: dynamic QR model; HDM: hybrid dynamic model; DSM: dynamic Smagorinsky model, and DMD: dynamic minimum-dissipation model.}
    \label{fig:channel4}
\end{figure}
The focus of this part is on the comparison of the static QR model with C = 0.024 and four dynamic models. Figure \ref{fig:channel4} illustrates the normalized mean streamwise velocity against wall distance in wall unit at $Re_\tau=180$. The grid and time step are the same as in the previous section. 

As shown in Figure \ref{fig:channel4}, all LES models are very close to the DNS results in the near-wall region, especially the dynamic QR model that is overlapping with DNS data. A closer inspection of this figure reveals that in the log wall range $y^+>30$, the dynamic QR model, static QR model, and the dynamic minimum-dissipation model are closer to DNS results. The dynamic Smagorinsky model, and hybrid dynamic model, however, are almost overlapping and apparently differ from the DNS results in the log wall region. Note that, all LES models result in a lower center-line mean velocity compared to DNS.

From the error quantification study, the dynamic QR model and static QR model perform more or less the same. The minor difference between the errors is insignificant to distinguish one model from another. Meanwhile, the dynamic Smagorinsky model and hybrid dynamic model perform equally less accurately, and the dynamic minimum-dissipation model has the highest error.

In conclusion, the investigation indicates that the static QR model is reliable. A properly chosen value of the model constant can provide very similar results to a dynamic model, at reduced computational cost. 

\subsubsection{Comparison of a symmetry-preserving discretization with standard OpenFOAM discretization}
\begin{figure}[htbp!]
    \centering
    \includegraphics[width=0.49\linewidth]{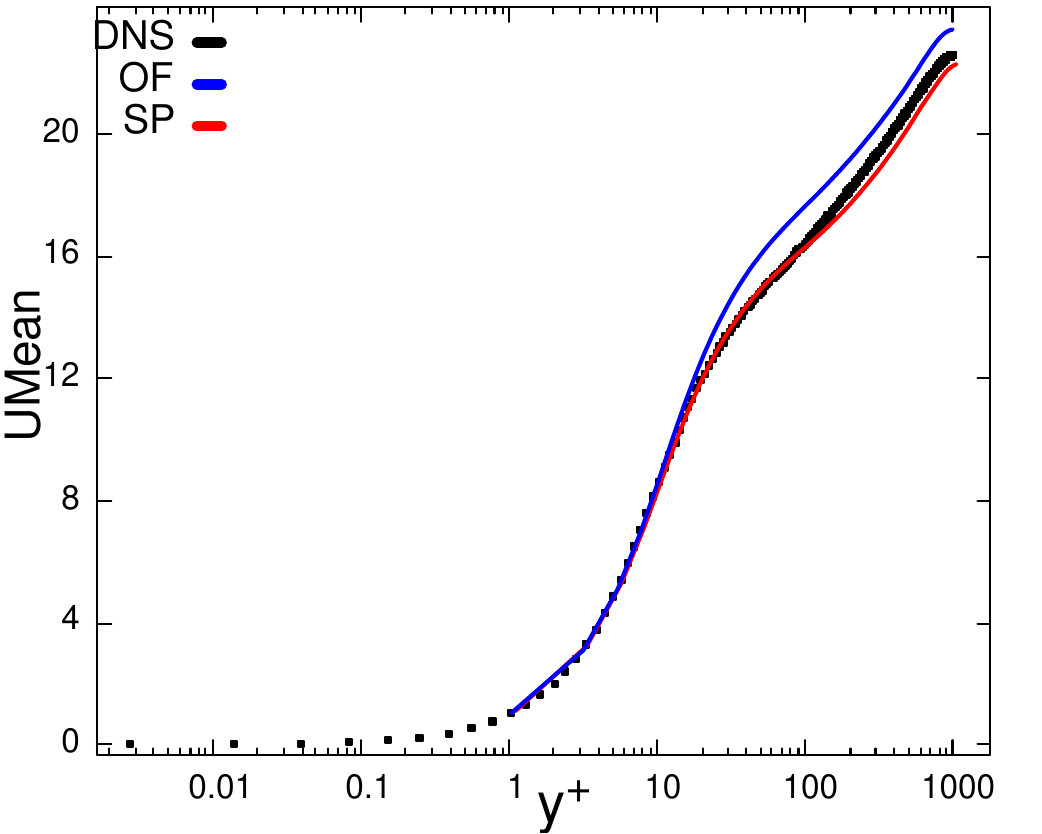}
    \includegraphics[width=0.49\linewidth]{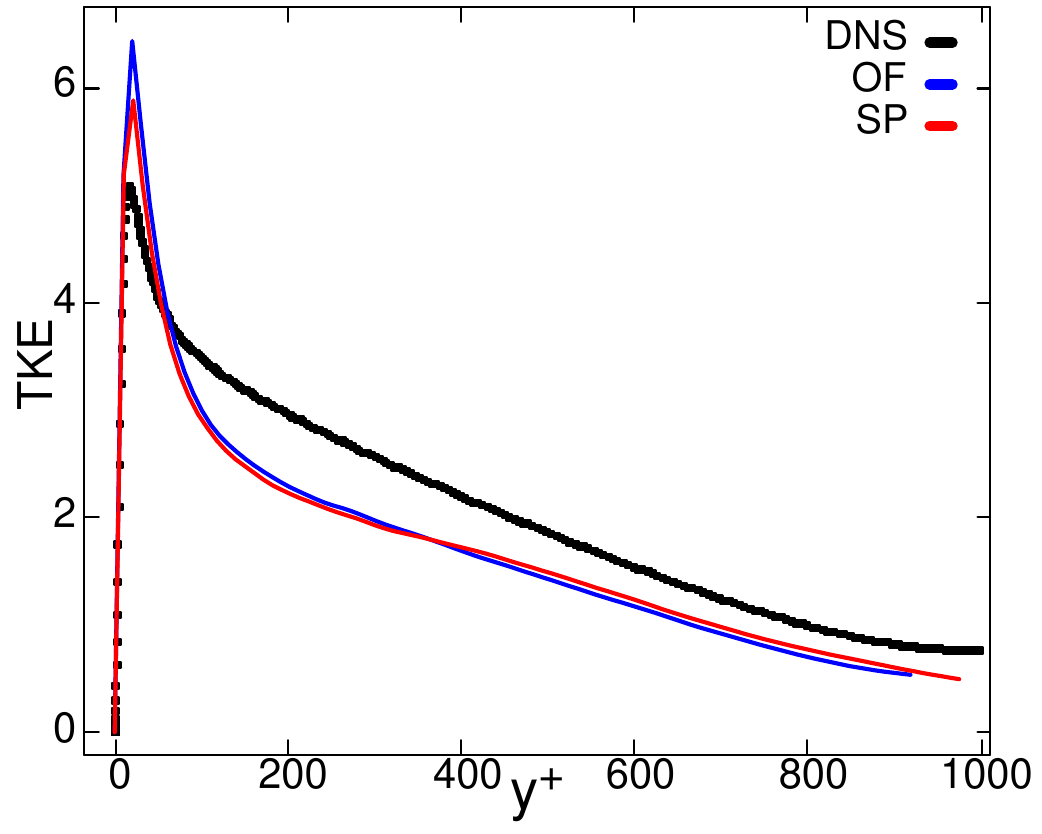}
    \includegraphics[width=0.65\linewidth]{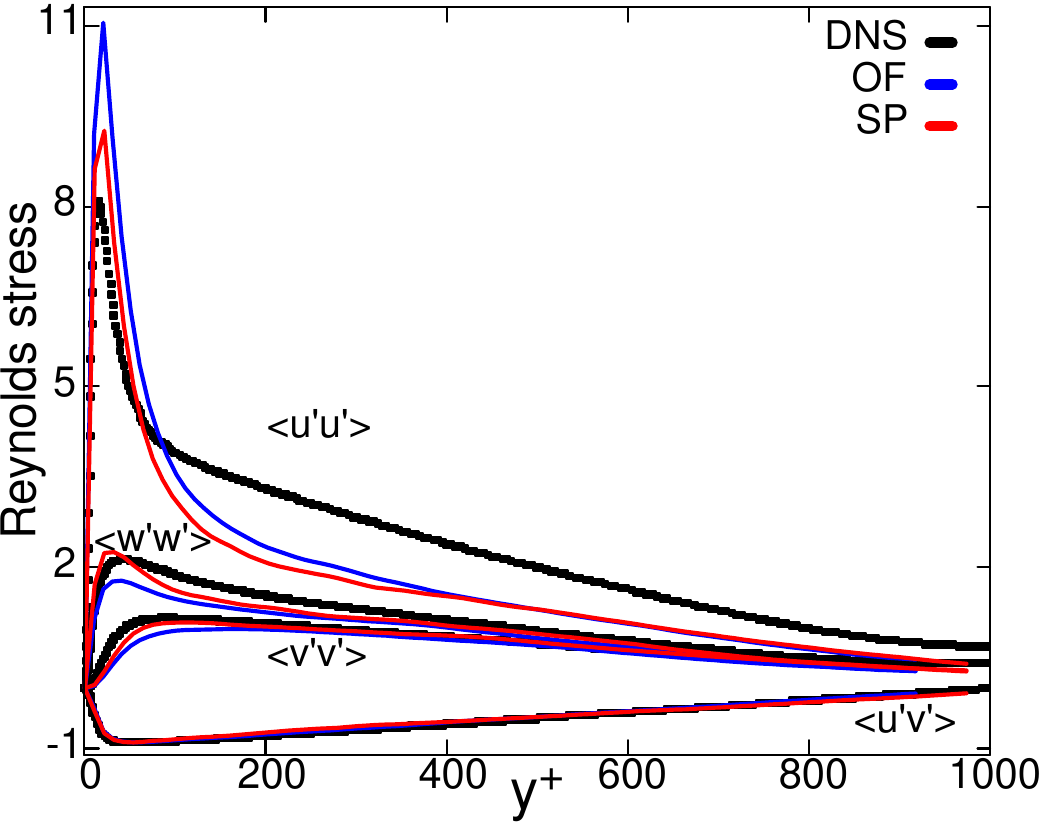}
    \caption{\label{fig:channelNum1}Comparison of reference DNS and QR simulations using OpenFOAM (OF) discretization schemes and symmetry-preserving schemes (SP) for the mean, RMS velocity profiles, and turbulent kinetic energy for fully developed turbulent channel flow at $Re_{\tau}$ = 1000.}
\end{figure}

In this section, we illustrate the effect of the numerical schemes by comparing the results from OpenFOAM's standard Gauss linear discretization with symmetry-preserving discretization schemes for a turbulence channel flow at $Re_\tau =1000$. Second-order implicit Crank-Nicolson schemes are used with symmetry-preserving discretization. Second-order backward temporal discretization is used with standard OpenFOAM simulations.  Two inner loops and one outer loop are used for each of these two simulations. The CFL number is limited to $0.8$ with the time-step $\Delta t^+ =80$.

The results are shown in Fig.\ref{fig:channelNum1}. 
For the mean streamwise velocity $u^+$, the values calculated by symmetry-preserving schemes are closer to the DNS reference. Especially in the region $y^+\geq 10$, the prediction of symmetry-preserving with the QR model matches the DNS exactly. 
For the velocity fluctuation in the streamwise direction $u'u'$, the symmetry-preserving method improves the prediction in the region $y^+ \leq 220$. In other regions of the computational domain, the two methods give similar results.

For the velocity fluctuation in the spanwise direction $w'w'$ and in the wall-normal direction $v'v'$, the symmetry-preserving discretization outperforms OpenFOAM discretization in the region $y^+ \leq 220$ and gives similar prediction in other regions. 
For the covariance of $u'$ and $v'$, the two methods give the same accurate prediction.  
For the turbulent kinetic energy $k$, the symmetry-preserving schemes predict relatively smaller over-predictions in the near-wall region.
It has been found that the contribution of the sub-grid scale model to the diffusive flux in both methods is much lower than the contribution of the molecular viscosity, $\nu_{sgs}\leq 0.2 \nu$. 

\section{Flow over periodic hills}
\begin{figure}[ht!]
\centerline{
 \includegraphics[trim=40 80 50 90, clip, width=0.98\linewidth]{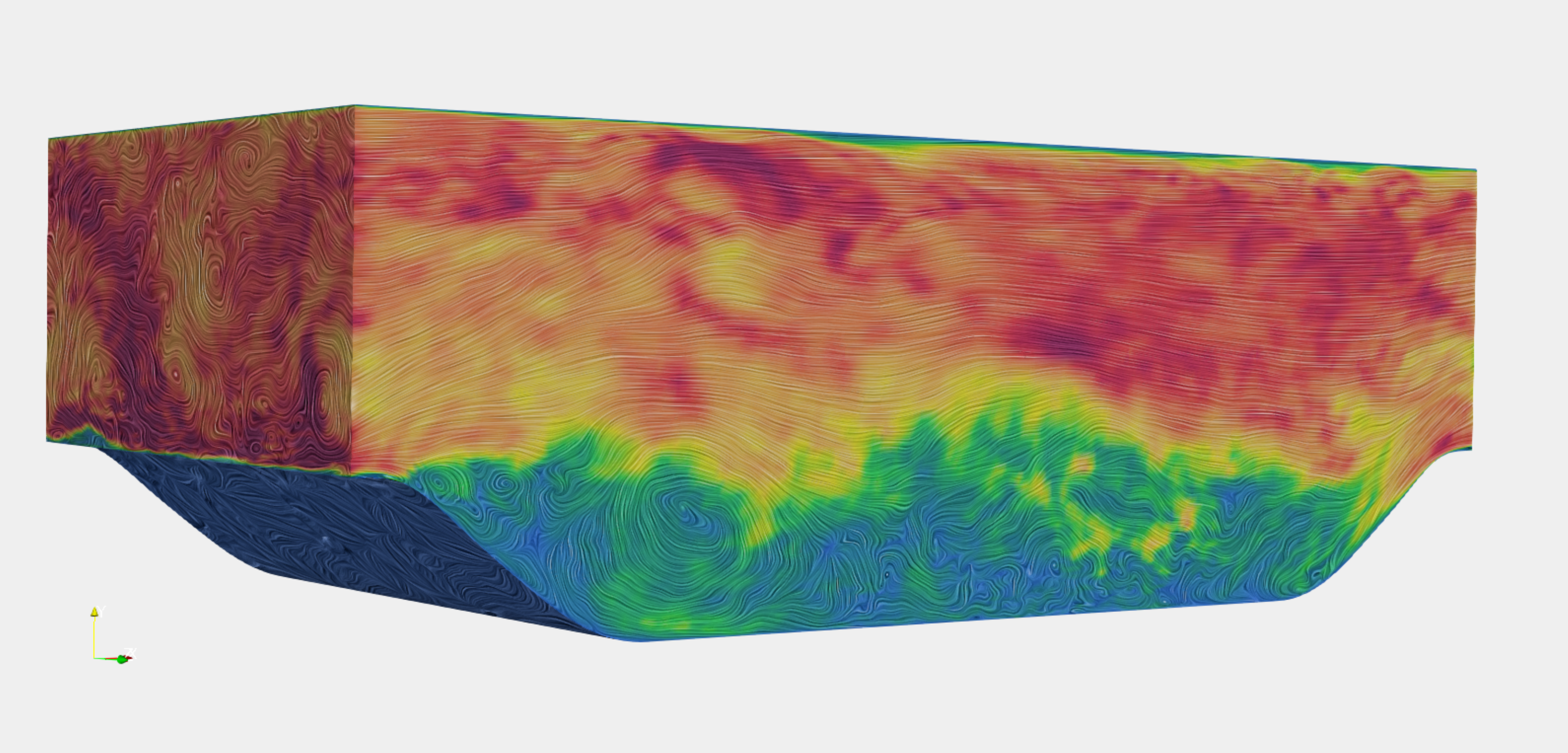}}
\caption{The side view of the geometry of the periodic hills. }
\label{fig:hill1}
\end{figure}
Flow separation from curved surfaces and subsequent reattachment is a flow phenomenon often appearing in engineering applications.  To assess the applicability of the proposed minimum-dissipation model in OpenFOAM to compute separated flows, simulations of three-dimensional flow over periodic hills at $Re=10595$ have been performed.

The geometry retains the shape of the hill defined by Mellen et al(2000).  
The flow is assumed to be periodic in the streamwise direction and thus periodic boundary conditions are applied.
Additionally, the flow is assumed to be homogeneous in the spanwise direction, and periodic boundary conditions are implemented accordingly. The simulations are conducted on a grid consisting of approximately 2.56 million points. The grid resolution near the wall is sufficient to resolve the viscous sublayer, as indicated by a $y^+$ value of approximately 0.17 at the closest grid points to the wall. Therefore, the no-slip boundary condition is employed at the wall.

\subsection{Cross-comparison of calculation from QR model using standard OpenFOAM discretization with literature data}
\subsubsection{Separation and Reattachment Lengths}
The separation and reattachment points are obtained at $Re=11230$, i.e. the bulk velocity $u_b=1.06$, using Gauss linear spatial discretion and Backward temporal discretization. The separation point is accurately determined by numerically solving the boundary layer equations under pressure-adverse conditions. The QR model predicts the separation point, where the wall shear stress reaches zero, to be approximately $x/H \approx 0.175$, which is smaller than the reference value of $x/H \approx 0.19$.
This discrepancy is reasonable since the separation point moves upstream with increasing Reynolds numbers. The separation point has a strong impact on the point of reattachment. The recirculation starts at $x/H \approx 0.27$ and ends on $x/H \approx 4.71$. The length of the main recirculation bubble is approximately 4.48. The reattachment position where the dividing streamline attaches to the wall again is $x/H = 4.71$, which is very close to the reference value of 4.69 (Rapp et al (2010)). 

\subsubsection{The effect of numerical schemes from standard OpenFOAM}
\begin{figure}[htbp!]
            \includegraphics[width=0.99\linewidth]{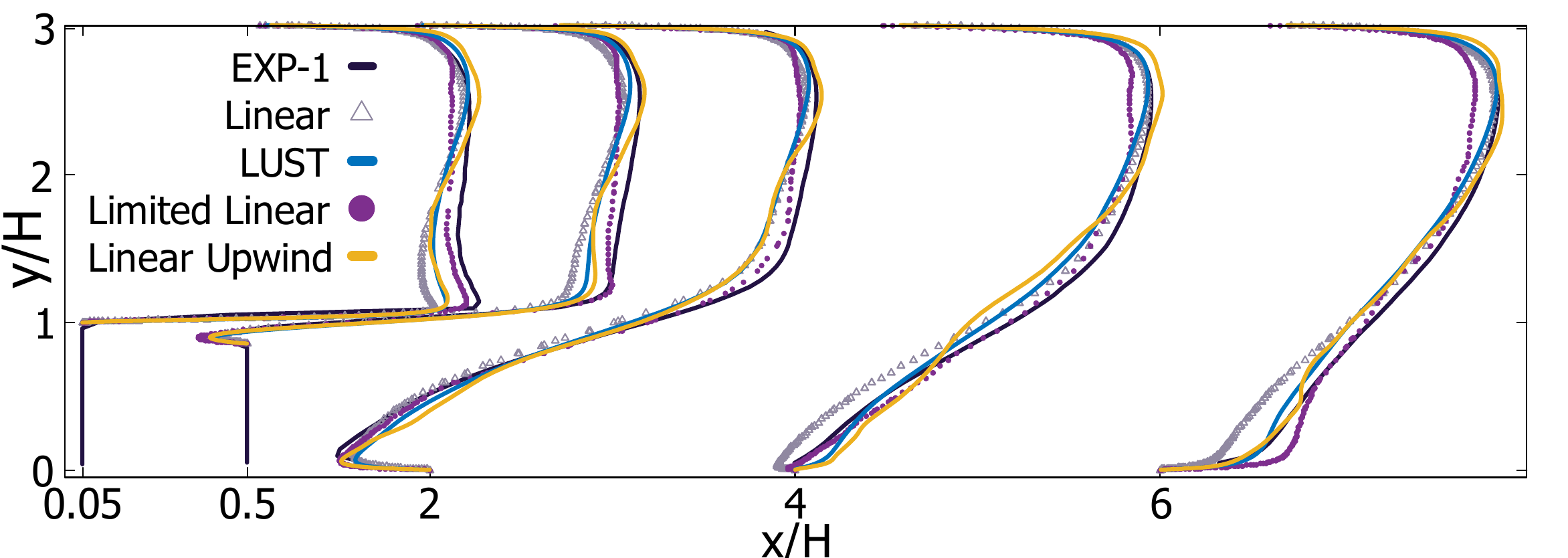}
            \captionsetup{font={footnotesize}}
            \caption{Streamwise velocity at five different positions with varying finite volume discretization schemes}
            \label{fig:hill3}
\end{figure}

In this section, different interpolation numerical schemes for solving divergence term and pressure gradient have been tested, including central difference, filtered and limited central difference, linear upwind blended with a central difference, least square, and detached eddy discretization schemes. The outcomes of these different numerical schemes, however, are not capable of predicting the mean velocity accurately, as shown in Figure \ref{fig:hill3}. Therefore, it is decided that the Reynolds number needs to be increased to match the reference data. In addition, the central difference schemes are selected anyway based on their decent performance in the previous validation cases, although minor differences between the simulations are insignificant to distinguish one technique from another.

\subsection{The effect of QR model constant}
\begin{figure}[htbp!]
    \centering
    \includegraphics[width=0.96\linewidth]{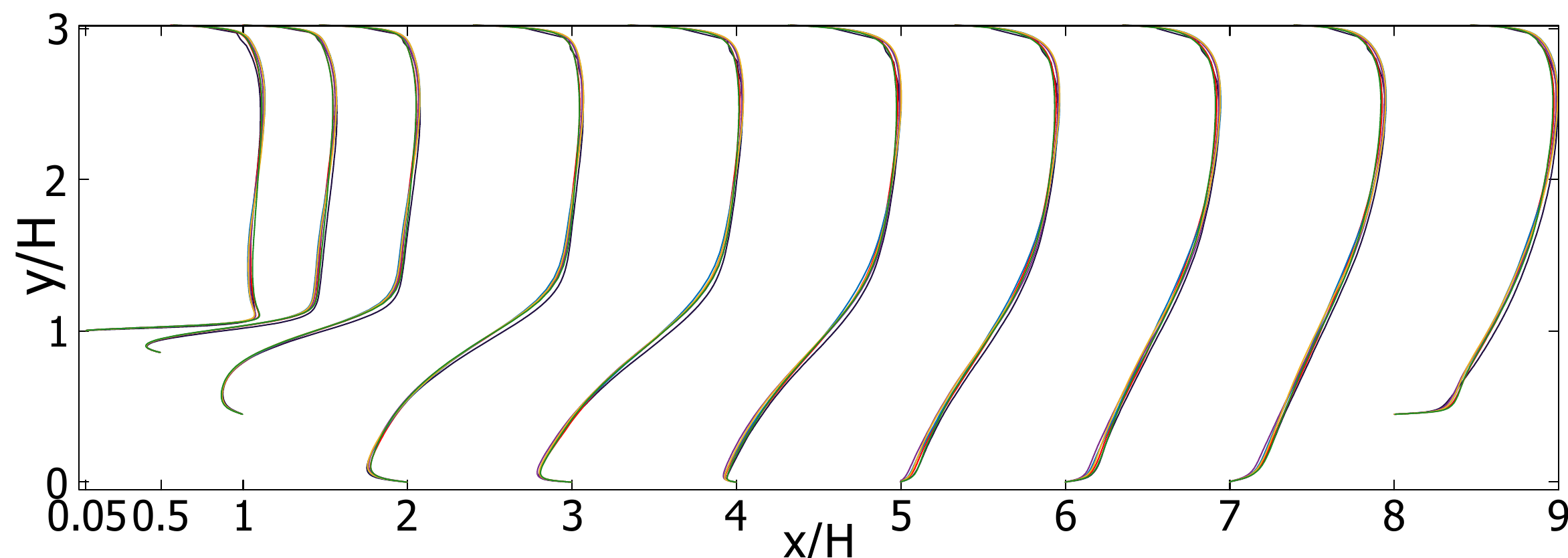}
    \includegraphics[width=0.96\linewidth]{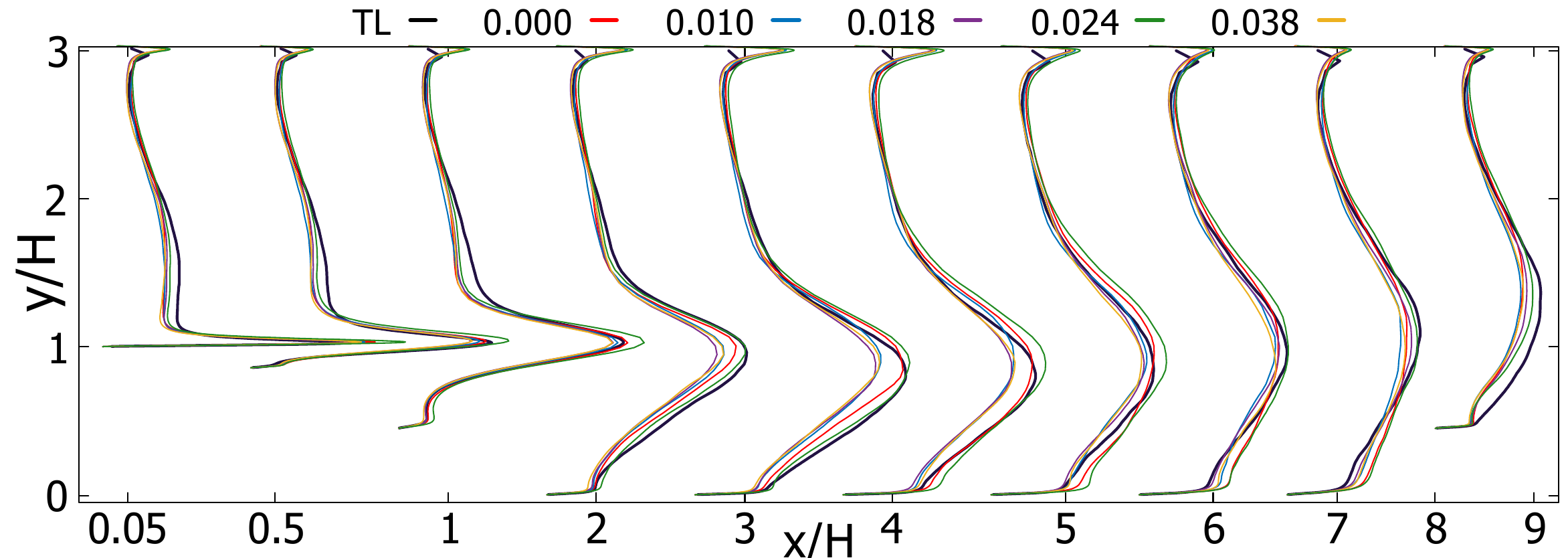}
    \caption{Comparison of the QR model constants using standard OpenFOAM discretization at ten different locations. Mean streamwise velocity (top) and Reynolds stress $u'u'$(bottom). TL refers to the experimental results from Temmerman and Leschziner (2001) }
    \label{fig:hill6}
\end{figure}To evaluate the performance of the minimum-dissipation model, simulations were conducted using different model constants ($C$) including no-model ($C=0$) and $C$ values of $0.01, 0.018, 0.24$, and $0.038$. Gauss linear interpolation was employed as the interpolation scheme, resulting in a central difference scheme on a uniform mesh. Temporal discretization was accomplished using a second-order implicit Backward approach.

Regarding the mean streamwise velocity, varying the model constant has no significant impact, as depicted in Figure \ref{fig:hill6}. Conversely, for the mean Reynolds stress in the streamwise direction ($u'u'$), the constant of $0.024$ closely approximates the reference within the region $1.1\leq y/H \leq 2$ and $x/H \leq 4$. This setting results in an overestimation of the peak velocity within the region $x/H \leq 1$ while approaching the experimental results at $x/H =2, 3, 6$, and $7$. On the other hand, $C$ values of $0.1, 0.18$, and $0.038$ underestimate the peak value of $u'u'$ across the entire computational domain but provide reliable predictions in the upper part of the domain ($y/H\geq 2$).

In conclusion, the model constant significantly influenced the root-mean-square variables, while yielding similar predictions for mean variables. A model constant of $0.024$ produced more accurate results, similar to the case of channel flow.

\subsection{Symmetry-preserving discretization compared to the standard OpenFOAM discretization schemes}
The symmetry-preserving discretization implemented in OpenFOAM is compared with the standard Gauss linear schemes in OpenFOAM and the experimental data obtained from Temmerman and Leschziner (2001). 
\begin{figure}[htbp!]
    \centering
    \includegraphics[width=0.96\linewidth]{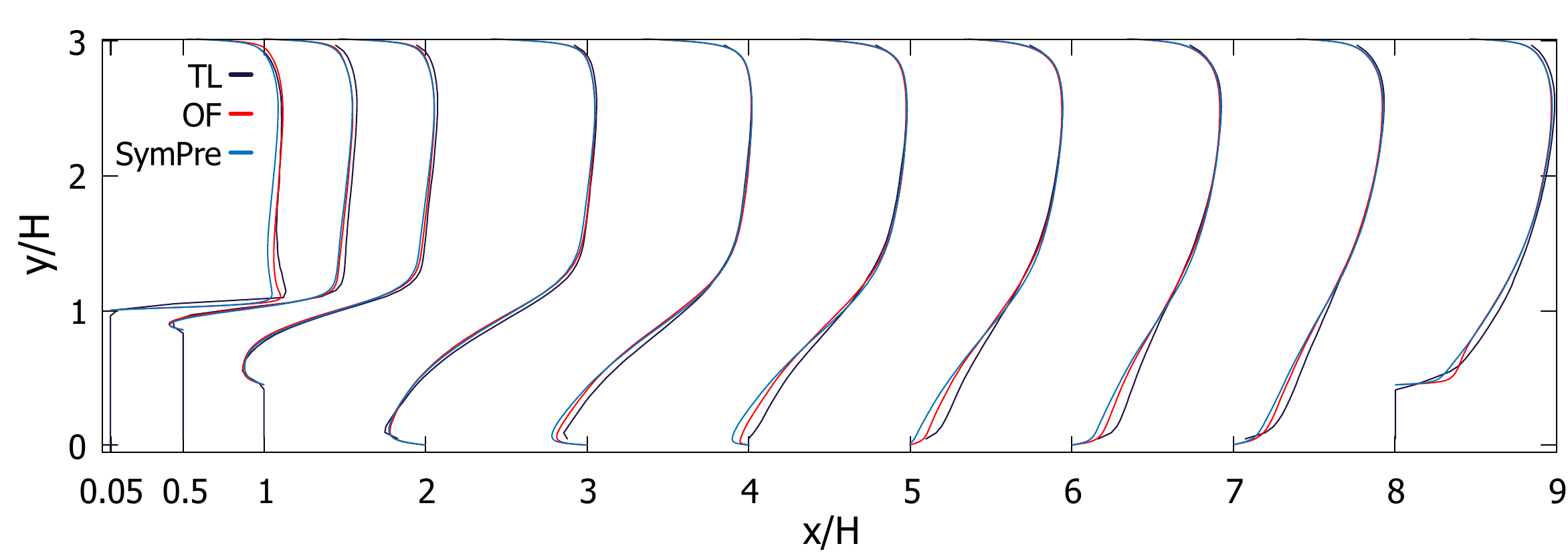}
    \includegraphics[width=0.96\linewidth]{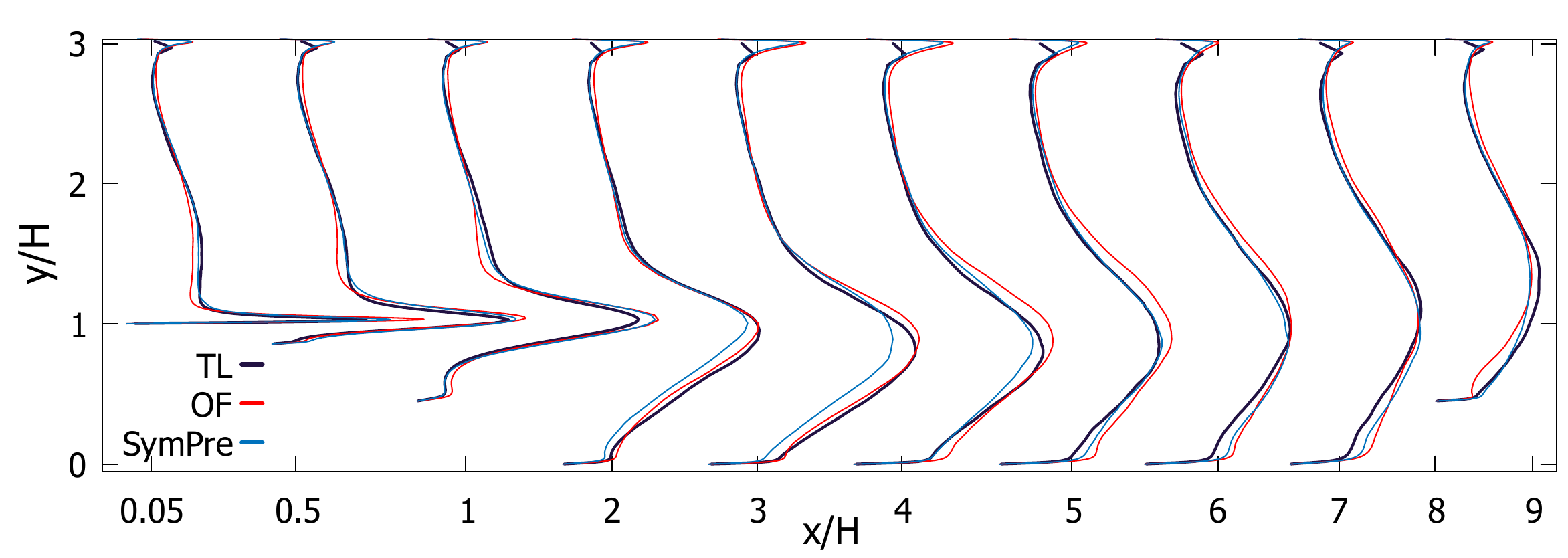}
    \includegraphics[width=0.48\linewidth]{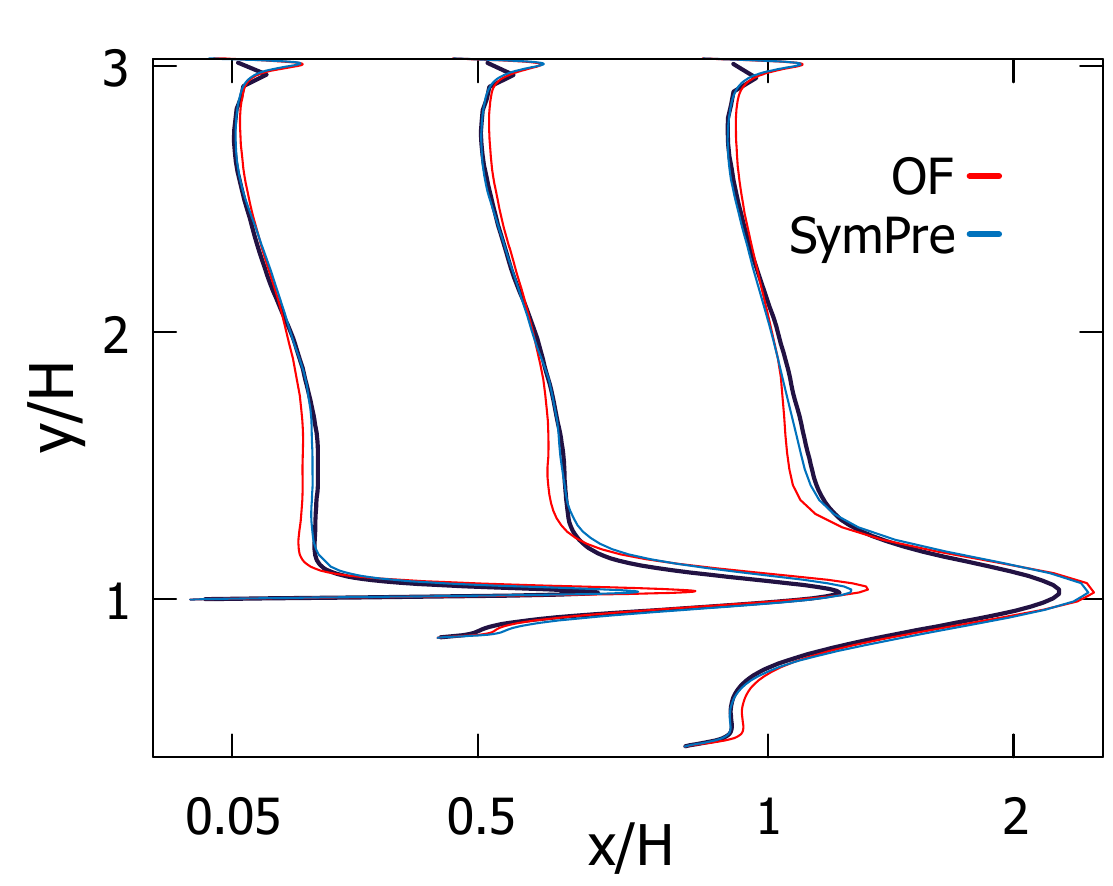}
    \includegraphics[width=0.48\linewidth]{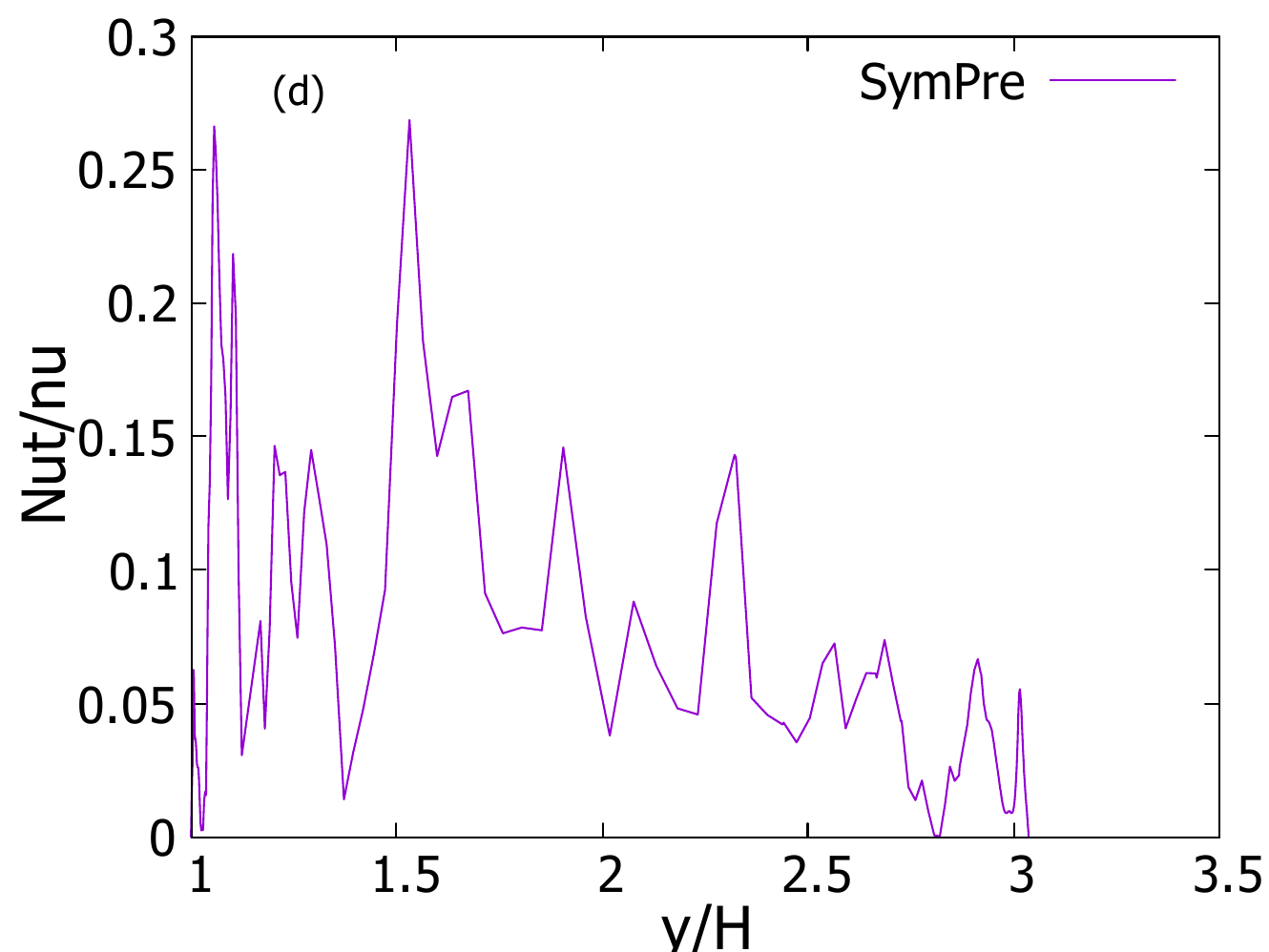}
    \caption{Comparison of the predictions by standard OpenFOAM discretization and symmetry-preserving discretization. Top: Mean streamwise velocity; middle: Averaged Reynolds stress in the streamwise direction $u'u'$ at ten different locations in the streamwise direction; Bottom left: Zoomed-in $u'u'$ at $x/H=0.05, x/H=0.5$ and $x/H=1$; (d): Eddy viscosity normalized by the fluid viscosity.}
    \label{fig:hill7}
\end{figure}
 The separation point predicted by the symmetry-preserving discretization is approximately $x/H \approx 0.175$, which is smaller than the reference value of $x/H \approx 0.19$. The recirculation starts at $x/H \approx 0.27$ and ends on $x/H \approx 5.02$. The length of the recirculation bubble is approximately $x/H\approx 4.7$. 

As we can see from Fig.\ref{fig:hill7}, the mean velocity predicted by the two discretization schemes is consistent in the upper part ($y/H> 1$) of the computational domain, where the structure is relatively simple, and no hill is present. On the bottom part ($y/H <1$), two simulations again consistent in the upstream region $x/H=0.05, 0.5, 1$ and $2$. The underprediction appears in both simulations at $x/H=3$ until the end of the domain.  

The Reynolds stress in the streamwise direction $u'u'$ in the middle of Fig.\ref{fig:hill7} shows the underprediction and overprediction at different locations. The trends are clearer if the upstream region is zoomed in, as shown in the bottom figure. In the first three locations, i.e. $x/H=0.05, 0.5$, and $1$, the standard OpenFOAM discretization underestimate the $u'u'$ at the middle ($1<y/H<2$), overestimates the peak value ($y/H=1$) and the $u'u'$ near the wall. Meaning the acceleration predicted by standard OpenFOAM is more intense in the shear layer at the hill crest. Fig.(d) in Fig.\ref{fig:hill7} shows the eddy viscosity normalized by the fluid viscosity in the spanwise direction, from which we can see the model contribution $\nu_t/\nu$ is below 0.3.

To sum up, when simulating periodic hills, the minimum-dissipation model, along with standard OpenFOAM discretization schemes and symmetry-preserving schemes, provides dependable results while significantly reducing computational expenses. The symmetry-preserving discretization yields more accurate outcomes in certain areas of the computational domain. Komen et al (2021) found that the numerical dissipation introduced by standard OpenFOAM discretization exceeds the contribution of the large-eddy model. Therefore, combining the symmetry-preserving discretization with the QR model is advantageous and dependable. 

\section{Flow over a circular cylinder }
\subsection{The influence of discretization scheme}
To study the influence of finite volume methods for the convective fluxes and the pressure gradient on the turbulence behaviors, the simulation with pure central difference schemes (Run \RomanNumeralCaps{3}) and $25\%$ upwind-biased central difference (Run \RomanNumeralCaps{4}) methods while preserving the same mesh resolution of Run \RomanNumeralCaps{2} are performed.

\begin{figure}[htbp!]
    \centering
    \includegraphics[height=0.17\paperheight,width=0.8\linewidth]{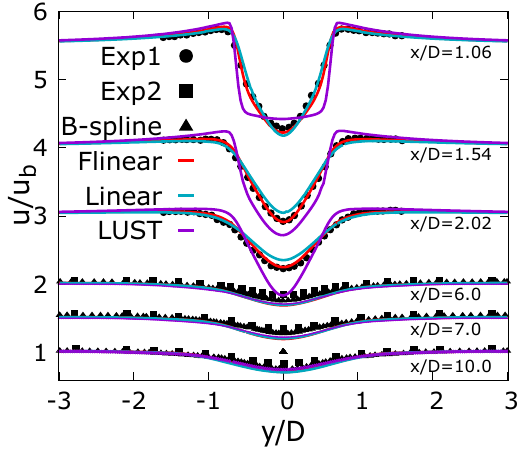}
    \caption{Mean velocity of the flow over cylinder at $Re_D=3900$ with domain size of $50D \times 30D \times \pi D$.}
    \label{fig:cylinder2}
\end{figure}

The mean streamwise velocity obtained from three numerical schemes differs in the near wake region ($x/D < 2.02$) as shown in Fig. \ref{fig:cylinder2}. Especially the upwind-blended scheme (Run \RomanNumeralCaps{4}) calculates U-shape mean velocity at $x/D = 1.06$, then develops a much lower V-shape profile further at $x/D = 1.54$ and $ 2.02$, comparing to other two schemes. 
The central difference simulations point out that the transition to turbulence in the separating shear layers occurs closer to the cylinder and leads to the development of the V-shape profile and shorter vortex formation region. As a result, the shear layers are shorter and the recirculation region is smaller. 
In the downstream location, it is found that varying the numeric schemes has no apparent effect on the mean velocity.  
\begin{figure}[htbp!]
    \includegraphics[width=0.49\linewidth]{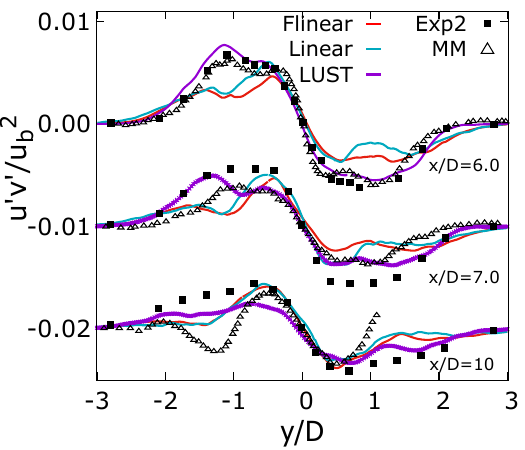}
    \includegraphics[width=0.49\linewidth]{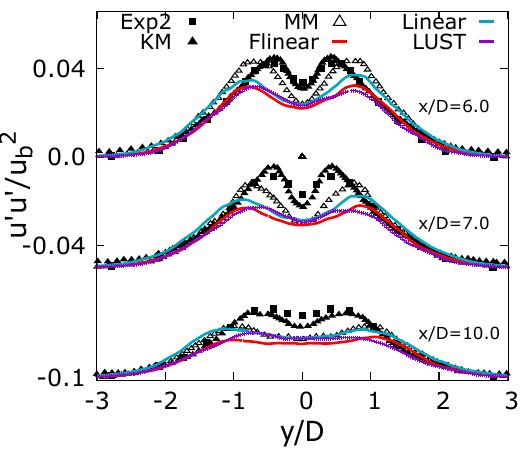}
    \caption{The turbulence fluctuations in the downstream.  Fliner: Filtered central difference; Linear: pure central difference; LUST: upwind-biased; the references are denoted by the symbols.}
    \label{fig:cylinder3}
\end{figure}

For cross-section Reynolds stress $v'v'$ at the position of $x/D=1.54$, the upwind-blended scheme (Run \RomanNumeralCaps{4}) predicts lower quantities to a large extent. The pure central difference scheme (Run \RomanNumeralCaps{3}) calculates values that are too large. In addition, the filtered central difference (Run \RomanNumeralCaps{2}) is in accordance with Breuer's simulation (Breuer (1998)) which used the Smagorinsky sub-grid model conjugated with a central difference, but both Run \RomanNumeralCaps{2} and Breuer's simulations overestimate $v'v'$ in comparison with the experiments of Lourenco and Shih. 

As for the streamwise Reynolds stress $u'u'$ shown in Figure \ref{fig:cylinder3}, the central difference (Run \RomanNumeralCaps{3}) improves the velocity fluctuation $u'u'$ to a small extent everywhere in the downstream region.  However, the minor differences between Run \RomanNumeralCaps{2} and Run \RomanNumeralCaps{3} are insignificant to distinguish one technique from another. 

The mean velocity at the central line shows the solutions of filtered central difference (Run \RomanNumeralCaps{2}) match the experimental data (Lourenco (1994)) very well in the near wake region, furthermore, the recirculation length of $1.189$ is in good agreement with the experimental (Lourenco (1994)) value of $1.18$. The upwind-blended scheme (Run \RomanNumeralCaps{4}), however, calculates a conspicuously long region of recirculation, but then shows good agreement of center line streamwise velocity further downstream, at $x/D > 7$, comparing with B-spline simulation and the hot-wire measurements of Ong and Wallace.

\section{ Conclusions } 

The results of channel flow mainly demonstrate the static QR model predicts equally accurate results than the dynamic models while reducing the computational cost, the model constant of $C=0.024$ gives the most accurate prediction, and the contribution of the sub-grid model decreases with the increase of the mesh resolution and becomes very small (less than 0.2 molecular viscosity) when the fine mesh is used. Furthermore, the QR model is able to predict the mean and invariance of turbulence accurately up to $Re_\tau = 2000$ without a wall damping function. The symmetry-preserving discretization outperforms the standard OpenFOAM discretization at $Re_\tau=1000$. The results for the flow over a cylinder show that mean velocity, drag coefficient, and lift coefficient are in good agreement with the experimental data and the central difference schemes conjugated with the QR model predict better results.
The various comparisons carried out for flows over periodic hills demonstrate the need to use central difference schemes in OpenFOAM in combination with the minimum dissipation model. The model constant of $C=0.024$ is again the best one for this case.


\Acknowledgments
The investigations presented in this paper have been obtained within the CSC-RUG joint project.

\begin{References}
    \item Breuer, M. (1998), Large eddy simulation of the subcritical flow past a circular cylinder: numerical and modeling aspects, \textit{Int J Numer Methods Fluids}, Vol. 28(9), pp. 1281–1302.
    \item Hopman, J., and Edo, F. (2023), Last accessed 13 June 2023. https://github.com/janneshopman/RKSymFoam.
    \item Komen, E. M., Hopman, J. A., Frederix, E., Trias, F. X., and Verstappen, R. W. (2021), A symmetry preserving second-order time-accurate piso-based method. \textit{Computers $\&$ Fluids}, Vol. 225, pp. 104979.
    \item Lourenco, L. (1994), Characteristics of the plate turbulent near wake of a circular cylinder. a particle image velocimetry study. \textit{In Unpublished, results taken from Beaudan and Moin}.
    \item Mellen, C., Frohlich, J., and Rodi, W. (2000), Large eddy simulation of the flow over periodic hills. IMACS Paper 21–25.
    \item Rapp, C., Pfleger, F., and Manhart, M. (2010), New experimental results for a les benchmark case.DLES Paper 69-74.
    \item Rozema, W., Bae, H. J., Moin, P., and Verstappen, R. (2015), Minimum-dissipation models for large eddy simulation. \textit{Phys. Fluids}, Vol. 27(8), pp. 085107.
    \item Temmerman, L., and Leschziner, M. A. (2001), Large eddy simulation of separated flow in a streamwise periodic channel constriction. Turbulence and Shear Flow Phenomena Paper.
    \item Trias, F., Lehmkuhl, O., Oliva, A., Perez-Segarra, C., and Verstappen, R. (2014). Symmetry-preserving discretization of navier–stokes equations on collocated unstructured grids. \textit{J. Comput. Phys.}, Vol. 258, pp. 246–267.
    \item Verstappen, R. (2011), When does eddy viscosity damp subfilter scales sufficiently? \textit{J Sci Comput}, Vol. 49(1), pp. 94–110.
    \item Verstappen, R. (2018), How much eddy dissipation is needed to counterbalance the nonlinear production of small, unresolved scales in a large-eddy simulation of turbulence? \textit{Computers $\&$ Fluids}, Vol. 176, pp. 276–284.
\end{References}

\end{document}